\documentclass[useAMS,usenatbib]{mn2e}

\usepackage{epsfig}
\usepackage{graphicx}
\usepackage{times}
\usepackage{natbib}

\newif\ifAMStwofonts
\AMStwofontstrue

\newcommand{\be}{\begin{equation}}
\newcommand{\ee}{\end{equation}}
\newcommand{\ba}{\begin{eqnarray}}
\newcommand{\ea}{\end{eqnarray}}
\newcommand{\brr}{\begin{array}}
\newcommand{\err}{\end{array}}
\newcommand{\bc}{\begin{center}}
\newcommand{\ec}{\end{center}}
\newcommand{\hm}{\,h^{-1}{\rm Mpc}}
\newcommand{\hk}{\,h^{-1}{\rm kpc}}
\newcommand{\msun}{\,h^{-1}{\rm M}_\odot}

\newcommand{\fl}{\,{\rm erg\,s^{-1}cm^{-2}}}

\newcommand{\lum}{\,{\rm erg\,s^{-1}}}
\newcommand{\vel}{\,{\rm km\,s^{-1}}}

\newcommand{\gadget}{{\small{GADGET-2}}}

\newcommand{\mincir}{\raise
  -2.truept\hbox{\rlap{\hbox{$\sim$}}\raise5.truept \hbox{$<$}\ }}
\newcommand{\magcir}{\raise
  -2.truept\hbox{\rlap{\hbox{$\sim$}}\raise5.truept \hbox{$>$}\ }}
\newcommand{\siml}{\raise
  -2.truept\hbox{\rlap{\hbox{$\sim$}}\raise5.truept \hbox{$<$}\ }}
\newcommand{\simg}{\raise
  -2.truept\hbox{\rlap{\hbox{$\sim$}}\raise5.truept \hbox{$>$}\ }}

\title[Simulating a proto-cluster] 
      {Simulating the formation of a proto--cluster at $z\sim 2$} 
\author[A. Saro, et al.]
       {A. Saro$^{1,2}$, S. Borgani$^{1,2,3}$, L. Tornatore$^{1,2,3}$, 
         G. De Lucia$^4$, K. Dolag$^4$ \& G. Murante$^{3,5}$
         \\
         $^1$ Dipartimento di Astronomia dell'Universit\`a di Trieste, via
         Tiepolo 11, I-34131 Trieste, Italy
         (saro,borgani,tornator@oats.inaf.it)\\ 
         $^2$ INFN -- National Institute for Nuclear Physics, Trieste,
         Italy\\ 
         $^3$ INAF, Osservatorio Astronomico di Trieste, via Tiepolo 11,
         I-34131 Trieste, Italy\\
         $^4$ Max-Planck-Institut f\"ur Astrophysik, Karl-Schwarzschild-Str. 1,
         D-85748 Garching bei M\"unchen, Germany 
         (gdelucia,kdolag@mpa-garching.mpg.de)\\
         $^5$ INAF, Osservatorio Astronomico di Torino, Strada Osservatorio 20,
         I-10025 Pino Torinese, Italy (murante@oato.inaf.it)
       }
\begin{document}

\date{Accepted ???. Received ???; in original form ???}


\maketitle

\label{firstpage}

\begin{abstract}
  We present results from two high--resolution hydrodynamical
  simulations of proto--cluster regions at $z\simeq 2.1$. The
  simulations have been compared to observational results for the
  so-called Spiderweb galaxy system, the core of a putative
  proto--cluster region at $z=2.16$, found around a radio galaxy. The
  simulated regions have been chosen so as to form a poor cluster with
  $M_{200} \simeq 10^{14} \msun$ (C1) and a rich cluster with $M_{200}
  \simeq 2\times 10^{15} \msun$ (C2) at $z=0$. The simulated
  proto-clusters show evidence of ongoing assembly of a dominating
  central galaxy. The stellar mass of the brightest cluster galaxy
  (BCG) of the C2 system is in excess with respect to observational
  estimates for the Spiderweb galaxy, with a total star formation
    rate which is also larger than indicated by observations. We
  find that the projected velocities of galaxies in the C2 cluster are
  consistent with observations, while those measured for the poorer
  cluster C1 are too low compared to the observed velocities. We argue
  that the Spiderweb complex resemble the high--redshift progenitor of
  a rich galaxy cluster. Our results indicate that the included
    supernovae feedback is not enough to suppress star formation in
    these systems, supporting the need of introducing AGN feedback.
  According to our simulations, a diffuse atmosphere of hot gas in
  hydrostatic equilibrium should already be present at this redshift,
  and enriched at a level comparable to that of nearby galaxy
  clusters. The presence of this gas should be detectable with future
  deep X--ray observations.
\end{abstract}


\section{Introduction} \label{sec:intro}

Within the standard cosmological model of structure formation, galaxy clusters
trace regions where the hierarchical build up of galaxies and their interaction
with the inter-galactic medium (IGM) proceed at a somewhat faster rate compared
to regions of the Universe with `average' density. Finding and observing
high--redshift progenitors of galaxy clusters can provide invaluable
information on the processes which led to their formation and evolution.  The
most distant clusters to date have been identified out to $z\mincir 1.5$ from
deep X--ray and infra-red observations \citep[e.g.][and references
  therein]{2005ApJ...623L..85M,2008arXiv0804.4798E,2005ApJ...634L.129S}. A
powerful technique, that extends these studies to higher redshift, is to search
for overdensities of emission line galaxies in the neighbourhood of luminous
high-redshift radio galaxies \citep[e.g.][]{1998ApJ...504..139P}. Although
these regions are expected not to trace virialized clusters, they likely
identify the progenitors of present day clusters, offering an unique
opportunity to study the evolutionary processes which determine their
observational properties. Recently, a number of observations of distant
putative proto-clusters, like the one associated with the so--called
``Spiderweb Galaxy'' at $z=2.16$ (\citealt{2006ApJ...650L..29M}, M06
hereafter), have demonstrated that these regions are characterised by intense
dynamical and star formation activity (e.g. \citealt{2000A&A...361L..25P}, M06,
\citealt{2007A&A...461..823V,2008ApJ...673..143O}, and references therein).

On the theoretical side, modern cosmological hydrodynamical simulations are now
reaching good enough resolution, and include detailed treatments of a number of
astrophysical processes, to provide an accurate description of the assembly of
galaxy clusters. These simulations are now able to reproduce the basic
properties observed for the bulk of the cluster galaxy population at low
redshift, while generally predicting too massive and too blue BGCs, due to the
absence on an efficient feedback mechanism that suppresses the star formation
activity in these galaxies at late times
\citep[e.g.,][]{2006MNRAS.373..397S,2008MNRAS.tmp..793R}.

In this Letter we present results from high--resolution simulations of
two proto--cluster regions at $z\sim 2$ that we compare to
observational results of the Spiderweb galaxy. The aim of this
analysis is to verify to what extent simulations of proto--cluster
regions in a standard cosmological scenario resemble the observed
properties of the Spiderweb complex.  Furthermore, we provide
predictions for the properties of the proto intra-cluster medium
(ICM), in view of future deep X--ray observations of high--redshift
proto-cluster regions.

\section{The simulated proto--clusters} 
\label{sec:sims}
We analyse simulations of two proto--cluster regions, both selected at
redshift $z\simeq 2.1$, which by $z=0$ will form a relatively poor
cluster (C1) and a rich cluster (C2). These two regions have been
extracted from two different lower resolution parent cosmological
boxes, and resimulated at higher force and mass resolution using the
Zoomed Initial Condition (ZIC) technique by
\cite{1997MNRAS.286..865T}. The parent simulations correspond to two
renditions of the $\Lambda$CDM cosmological model, with the same
values for the parameters $\Omega_m = 0.3$, $h_{100} =0.7$ and
$\Omega_b = 0.04$, and different values for the normalisation of the
power spectrum ($\sigma_8 = 0.8$ for C1 -
\citealt{2004MNRAS.348.1078B}, and $\sigma_8=0.9$ for C2 -
\citealt{2001MNRAS.328..669Y}, see also cluster g8 high resolution from
\citealt{2008arXiv0808.3401D}).

The Lagrangian regions where mass and force resolutions were increased
extend out to $\sim 10$ virial radii of the clusters at $z=0$. The
total number of high resolution DM particles in these regions is about
$7.8\times 10^6$ for C1 and $2.4\times 10^7$ for C2, with an initially
similar number of gas particles.  The basic characteristics of the
simulated clusters at $z=2.1$ are summarised in Table \ref{t:clus}. At
$z=0$, the masses of these clusters are $M_{200} \simeq1.6\times
10^{14} \msun$ and $1.8\times 10^{15} \msun$ for C1 and C2
respectively\footnote{We define the mass $M_{200}$ as the mass
  contained within the radius, $r_{200}$, which encompasses an average
  density of 200 times the critical density.}. The mass of each gas
particle is $m_{gas}\simeq 1.5\times 10^{7}\msun$ for C1, and
$2.8\times 10^{7}\msun$ for C2. The Plummer--equivalent softening
scale for the gravitational force is $\epsilon_{Pl}=2.1h^{-1}$ kpc for
C1 and $2.75\,h^{-1}$ kpc for C2, in physical units
\citep[e.g.][]{2006MNRAS.367.1641B}. We verified that, by suitably
rescaling particle velocities in the initial conditions of the C2
cluster so as to decrease $\sigma_8$ to 0.8, the virial mass of this
system decreases by about 30 per cent, both at $z=0$ and at $z=2.1$.

The simulations were carried out using the TreePM--SPH {\small
  GADGET-2} code \citep{2005MNRAS.364.1105S}, including the effective
star formation model by \cite{2003MNRAS.339..289S} and the chemical
enrichment model by \cite{2007MNRAS.382.1050T}. The effective model of
star formation assumes that dense star--forming gas particles have a
cold neutral and a hot ionised component, in pressure equilibrium. The
cold component represents the reservoir for star forming material.

Within the adopted stochastic scheme of star formation, each gas
particle can spawn three star particles, with mass $m_*\simeq
0.5\times 10^{7}\msun$ for C1, and $0.9\times 10^{7}\msun$ for C2. The
chemical enrichment model assumes a Salpeter shape
\citep{1955ApJ...121..161S} for the stellar initial mass function
(IMF), and uses the same yields and stellar lifetimes adopted in
\citet{2008MNRAS.386.1265F}.  The simulations also include the kinetic
feedback from galactic outflows introduced by
\cite{2003MNRAS.339..289S}. We assume $v_w=500\vel$ for the wind
velocity, and a mass upload rate equal to two times the local star
formation rate. With these choices, the kinetic energy of the outflows
is roughly equal to all the energy available from SNII for the adopted
IMF.

\begin{table} 
\centering
\caption{Characteristics of the two simulated proto-clusters at $z \simeq
  2$. Col.  2: mass contained within $r_{200}$ (in units of
  $10^{13}\msun$). Col.  3: $r_{200}$ (in physical $\hk$). Col. 3 and 4: total
  stellar mass (in units of $10^{12}\msun$), and number of identified galaxies
  within $r_{200}$. Col. 5: X--ray luminosity in the [0.5--2.0] keV
  observer-frame energy band (units of $10^{43} \fl$). Col. 6:
  spectroscopic--like temperature. Col. 7: emission weighted Iron metallicity
  within $r_{200}$.}
\begin{tabular}{lccccccc}
Cluster & $M_{200}$ & $r_{200}$ & $M_*$ & $N_{gal}$ & $L_X$ & $T_{sl}$ & $Z_{Fe}$ \\ 
\hline 
C1 & 2.9 & 234.7 & 1.2  & 491 & 3.6  & 1.7 & 0.37\\
C2 &21.9 & 452.6 & 6.6  & 1571 & 41.3 & 4.7 & 0.46\\
\end{tabular}
\label{t:clus}
\end{table}

Galaxies are identified by applying the SKID algorithm\footnote{{\tt
    http://www-hpcc.astro.washington.edu/tools/skid.html} }
\citep{2001PhDT........21S} to the distribution of star particles,
using the procedure described in \citet{2007MNRAS.377....2M}.  We
identify as ``bona fide'' galaxies only those SKID--groups containing
at least 32 star particles after the removal of unbound stars.  Each
star particle is treated as a single stellar population (SSP), with
formation redshift $z_f$, metallicity $Z$, and appropriate
IMF. Luminosities in different bands are computed using the
spectro-photometric code GALAXEV \citep{2003MNRAS.344.1000B}, as
explained in \cite{2006MNRAS.373..397S}. We have computed luminosities
of galaxies in the two HST ACS camera (observer frame) filters F475W
($g_{475}$) and F814W ($I_{814}$). All magnitudes given in the
following are AB magnitudes. The effect of dust attenuation is
included by adopting the two-component model by
\cite{2000ApJ...539..718C}, in a similiar way as described in
\cite{2007MNRAS.375....2D} (see also \citealt{Fontanot08}). We have
divided our analysed box into a grid of cubic cells of linear size
$A$. We have then computed the optical depth $\tau_V$, contributed by
each cell, as
\begin{eqnarray}
  \tau_V \,=\, \left( {Z_{gas}\over Z_\odot }\right)^{1.6} \left({N_{H}\over 2.1 \times
      10^{21}cm^{-2}}\right)\,,
\label{eq_tau_v}
\end{eqnarray}
where $N_H=M_{gas}/(1.4m_pA^2)$, $M_{gas}$ is the mass of cold gas within the 
cell, $Z_{gas}$ is the mean metallicity of this gas, and $m_p$ is the proton
mass. The resulting optical depth is much less than unity for the majority of
the lines of sight, while it is $\tau_V\gg 1$ in the regions of most intense
star formation, where $\tau_V$ reaches values of about a thousand.

The dust--attenuated luminosity contributed by a star particle, having age $t$
and coordinates $(x_i,y_i,z_i)$ at the position $(x_i,y_i,z_f)$ can be computed
as: 
\begin{equation}
L_{\lambda,z_f}=
L_{\lambda,z_i}\exp\left(-\sum_{j=i}^{f}\tau_{\lambda}(x_i,y_i,z_j)\right),
\label{eq_L}
\end{equation}
where $\tau_\lambda$ is the optical depth at the wavelength $\lambda$ and is
related to $\tau_V$ by the relation \be \tau_\lambda = \left\{
\begin{array}{ll}
\tau_V (\lambda/550 \,nm)^{-0.7}  & \textrm{if $t \le 10^7 yrs$} \\
\mu \, \tau_V (\lambda/ 550 \,nm)^{-0.7} & \textrm{if $t > 10^7 yrs$} \\
\end{array}
\right.
\label{eq_tau_l}
\ee
\noindent
where $\mu$ is drawn randomly from a Gaussian distribution with mean 0.3 and
width 0.2, truncated at 0.1 and 1 (\citealt{2007MNRAS.375....2D}, see also
\citealt{2004MNRAS.349..769K}).  

\section{Results}
\label{sec:res}

\begin{figure*}
\vspace{-0.3truecm}
  \centerline{ \hbox{ \psfig{file=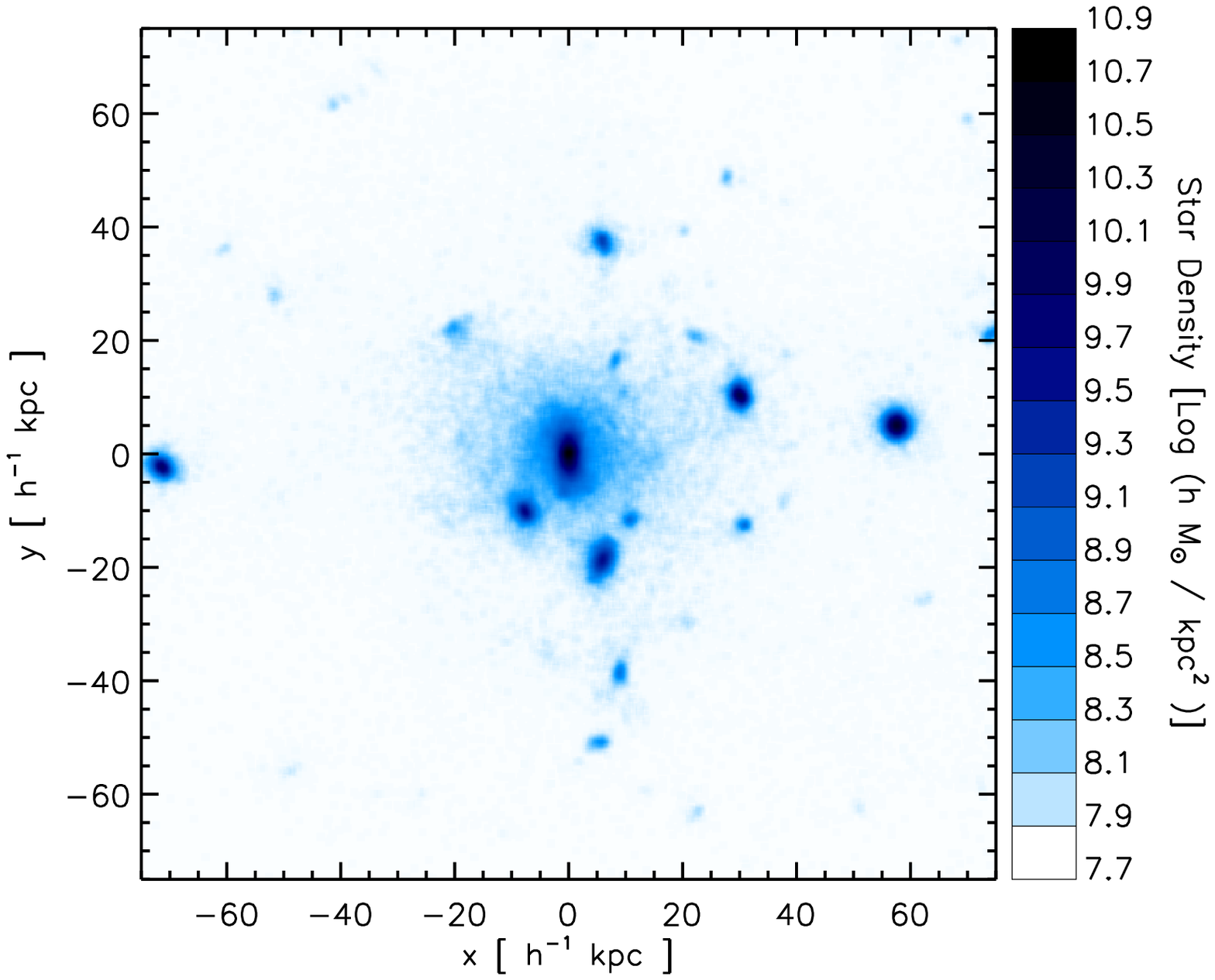,width=9.0cm}
      \psfig{file=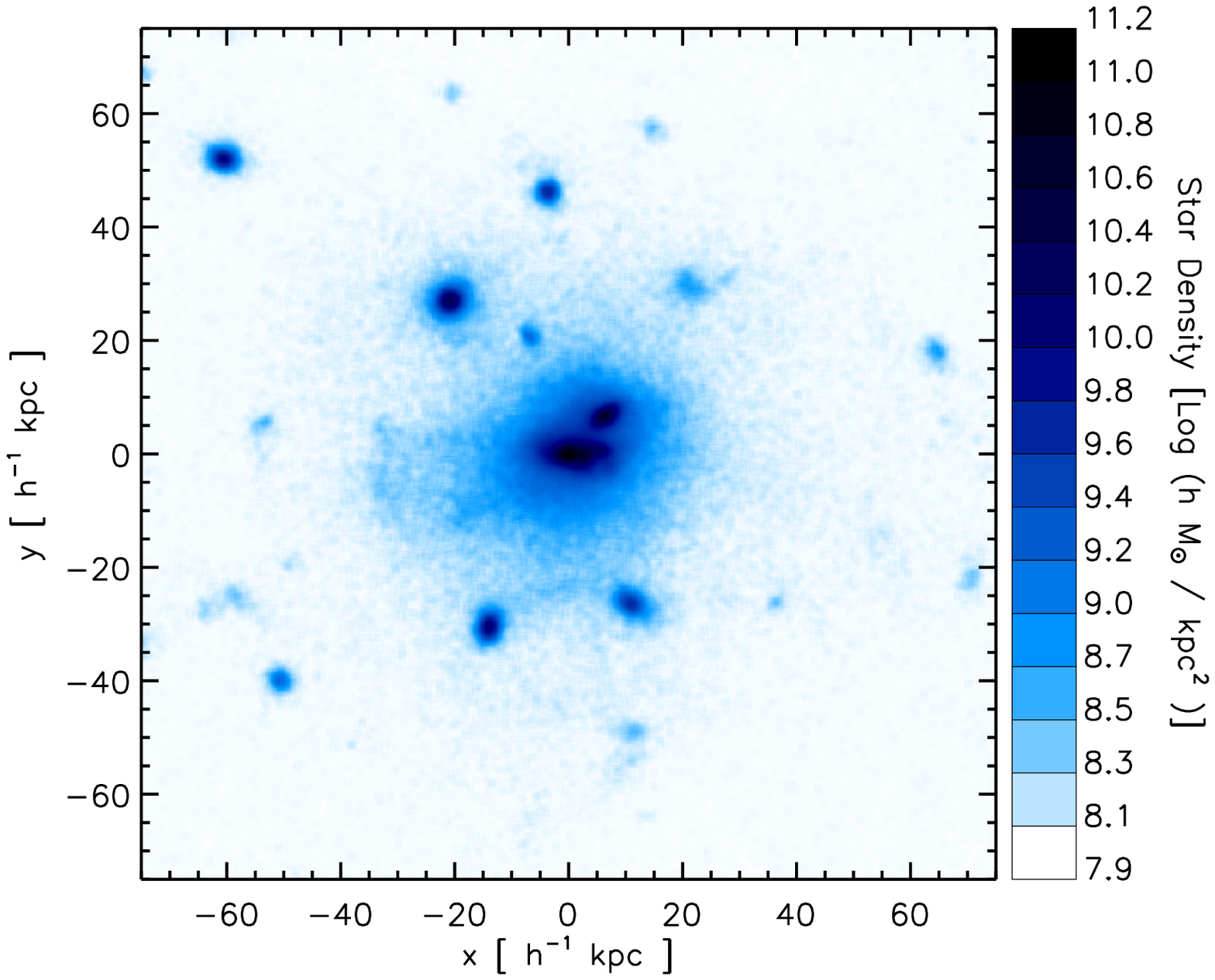,width=9.0cm} }}
\vspace{-0.5truecm}
  \caption{The projected stellar mass density for the C1 (left panel) and C2
    (right panel) clusters at $z \simeq 2.1$, within a region of $150\hk$ on a
    side, centred on the most massive galaxy.}
  \label{fi:Maps_star} 
\end{figure*}

\noindent{\bf Mass and luminosity of proto-cluster galaxies.}
Fig.~\ref{fi:Maps_star} shows the projected stellar mass density maps for C1
(left panel) and C2 (right panel). Simulated maps extends in projection along
the whole high-resolution region and are $150\,h^{-1}$kpc on a side, covering
an area which roughly corresponds to the physical size of the ACS images for
the assumed cosmology. The maps in Fig.~\ref{fi:Maps_star} are qualitatively
similar to observations of the Spiderweb complex (M06), in particular for the
C1 cluster, where the most massive galaxy is surrounded by several galaxies of
comparable mass in the process of merging to form a dominant central
object. The C2 cluster shows evidence of ongoing merging processes onto the
BCG, which exhibits a double peaked stellar mass density. In both clusters, a
dominant central galaxy is already in place.

Both clusters have a conspicuous population of inter--galactic stars. The
presence of a massive BCG and the visual impression of a significant ongoing
merging activity suggest that the merging processes associated to the BCG
assembly might be responsible (at least in part) for the generation of the
intra--cluster light
\citep[e.g.][]{2005MNRAS.358..949Z,2007MNRAS.377....2M,2008A&A...483..727P}.
Within the region shown, the diffuse stellar component contributes 16 and 30
per cent of the total stellar mass for C1 and C2 respectively. We also note the
presence of several elongated galaxies, similar to the tadpole galaxies
observed in the Spiderweb complex (M06).

The BCG of the C2 cluster has a stellar mass of $4.8\times 10^{12} \msun$,
about 6 times more massive that the BCG of C1, which is $7.9\times
10^{11}\msun$. The second most massive galaxy in the cluster C1 is about a
factor four less massive than the BCG, while the second most massive galaxy in
the cluster C2 is about a factor ten less massive than the corresponding BCG.
\cite{1998ApJ...504..139P} estimated the stellar mass of the radio-galaxy in
the Spiderweb complex from its $K$-band luminosity. They found a value of $\sim
10^{12} \msun$, comparable to the mass of the C1 BCG, but much less massive than
the C2 BCG. 

\begin{figure}
\vspace{-0.3truecm}
  \centerline{ \hbox{ \psfig{file=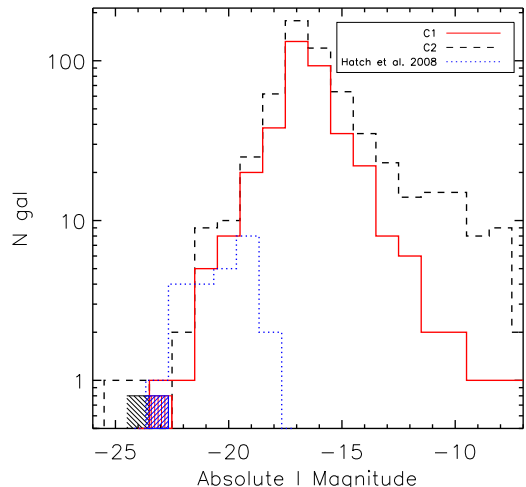,width=9.0cm}}}
\vspace{-0.5truecm}
\caption{Galaxy luminosity function within the region shown in Figure
  \ref{fi:Maps_star}, in the F814W($I_{814}$) ACS filter band. Results for the
  C1 and for the C2 simulated clusters are shown as solid and dashed histograms
  respectively. The dotted blue histogram shows the observed luminosity
  function (H08) within the same region. The shaded regions show the
    luminosity of the observed and of the simulated BCGs, with the same colour
    code described above.}
  \label{fi:LF} 
\end{figure}

\begin{figure}
\vspace{-0.3truecm}
  \centerline{\psfig{file=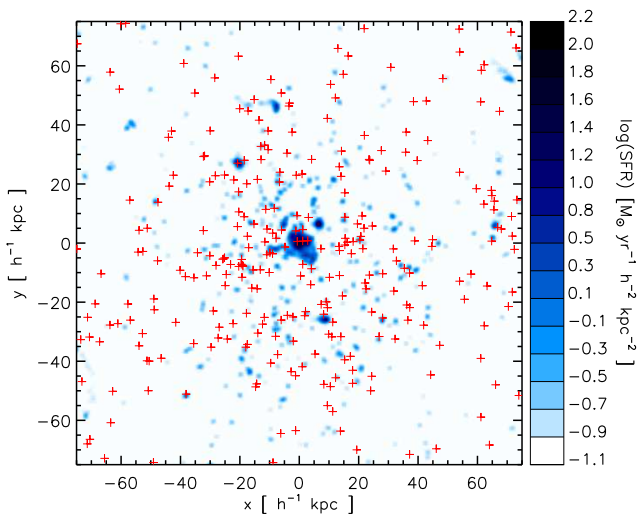,width=10.0cm} }  
\vspace{-0.5truecm}
  \caption{The map of star formation rate density for the C2 cluster, within
    the same region shown in Fig. 1. Red crosses mark the positions of the 50
    per cent most massive galaxies identified in projection within this
    region.}
  \label{fi:Maps_SFR} 
\end{figure}

Fig.~\ref{fi:LF} shows the luminosity functions (LFs) of the galaxies
identified in the two regions shown in Fig.~\ref{fi:Maps_star}, in the
F814W($I_{814}$) ACS filter band, including the dust correction
described in the previous section. Observational determination by
H08\footnote{The luminosity function published by H08 did not include
  the correction factor $2.5\times log_{10}(1+z)$ (R.A. Overzier,
  private communication). We have included this correction in the
  observed luminosity function plotted in Fig.~\ref{fi:LF}} are also
shown as a dotted histogram.  With the adopted dust correction, the
luminosity of the C1 BCG is comparable to the observed one, while that
of the C2 BCG is about one magnitude brighter than
observed. Interestingly, the brightest galaxy in the C2 cluster is not
the central galaxy, but a massive galaxy at the edge of the analyzed
box (it corresponds to the galaxy at the position $x \simeq -60$ and
$y \simeq 50$ in the right panel of Fig.~\ref{fi:Maps_star}). This
particular galaxy appears to be quite massive but not star forming
(see Fig.~\ref{fi:Maps_SFR}). It does not have a significant cold gas
reservoir and therefore is only weakly attenuated. The BCGs are
indicated by shaded regions in Fig.~\ref{fi:LF}, and appear to be in
reasonable agreement with observational measurements. This is in
apparent contradiction with the above mentioned evidence for a too
large mass of the C2 BCG. This can be explained by the very large star
formation rates that are associated to the C2 BCG (see below), which
can turn into an excessively large dust extinction. Therefore, having
reproduced the UV luminosity does not guarantee that we have correctly
described the star formation history of the BCG.

We note that the luminosity functions in our simulations rise steadily at
magnitudes fainter than the completeness limit of the observational data. These
fainter galaxies are expected to contribute to the budget of diffuse light by
an amount which depends of their surface brightness. Summing up the stellar
masses of galaxies fainter than $I_{814} = -19$ , our simulations predict that
the fraction of diffuse stellar component increases by 8 and 5 per
cent for the C1 and C2 cluster respectively.\\

\noindent{\bf Star formation:} Using the observed UV continuum, H08
traced the star formation pattern in the Spiderweb complex, finding
evidence of diffuse star formation, not associated to any detected
galaxy. We show in Fig.~\ref{fi:Maps_SFR} the star formation rate (SFR
hereafter) density map for C2 (results are qualitatively similar for
C1). Red crosses mark the positions of the half most massive galaxies
identified within the analysed region. The star formation in the
simulations is computed for each star--forming gas particle
\citep{2003MNRAS.339..289S}, so Fig.~\ref{fi:Maps_SFR} gives a
snapshot of the instantaneous SFR. The total SFR in the region
considered is $\sim600$ M$_\odot$/yrs for C1 and $\sim 1750$
M$_\odot$/yrs for C2. H08 suggest a total SFR for the Spiderweb system
of $\simeq130$ M$_\odot$/yrs within an area of 65 kpc$^2$ without any
dust correction. Assuming a minimum dust correction of $E(B-V)\simeq
0.1$, they estimate a lower limit of $ \simeq 325$ M$_\odot$/yrs,
which is reasonably close to the SFR measured for C1, but quite far
from our predictions for C2. We note that this excess of star
formation for the C2 BCG arises already at $z\sim 2$ and despite a
rather strong SN feedback (see Sec.~2).

Comparing the positions of the galaxies with the distribution of the
star formation, it is apparent that they do not always coincide.  With
the exclusion of the BCGs, which dominate the total SFR in both
clusters contributing from $55$ (C1) to $67$ (C2) per cent of the
total SFR, most of the massive galaxies are not strongly star
forming. The total SFR of identified galaxies accounts for 90 per cent
and 81 per cent of the total SFR for the clusters C1 and C2
respectively. The remaining diffuse star formation appears to take
place in clumps, which mainly correspond to halos falling below the
SKID galaxy identification threshold, and only in a few cases are
associated to cold stripped gas. Therefore, as long as the barely
resolved halos are identified with galaxies, our simulations do not
predict any significant diffuse star formation.\\

\begin{figure}
\vspace{-0.3truecm}
  \centerline{ \hbox{ \psfig{file=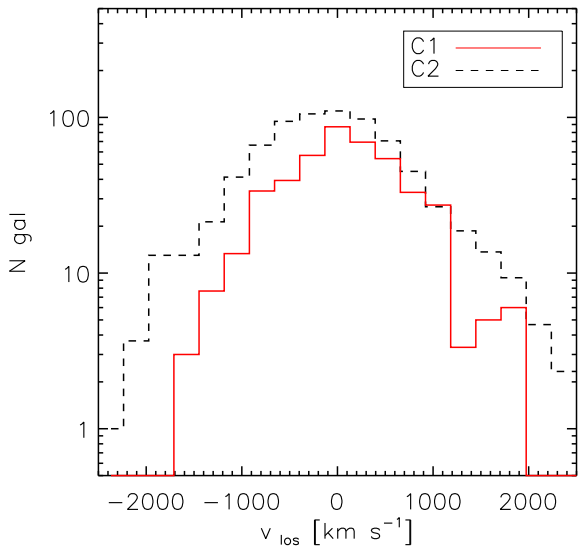,width=10.0cm}}}
\vspace{-0.5truecm}
\caption{Histogram of line-of-sight galaxy velocities the C1 (solid line) and
  C2 (dashed line) clusters. Each histogram shows the average over three
  independent projections of a square region of $150 \hk$ on a side.}
  \label{fi:Vel} 
\end{figure}

\noindent{\bf Are the proto-clusters in virial equilibrium?}  M06 measured
line-of-sight velocities for a small number of galaxies and found fairly large
values relative to the Spiderweb galaxy, up to almost 2000 $\vel$ for one
galaxy. Although the rather limited statistics prevent an accurate virial
analysis of the system, these values hint at a rather large virialized mass.
By computing the one-dimensional velocity dispersion ($\sigma_v$) for the fifty
most massive galaxies contained within $r_{200}$ (i.e. those which more likely
will have a redshift measurement in the Spiderweb complex), we find
$\sigma_v=481\vel$ for C1 and $916\vel$ for C2. These velocities are found to
be only 6--8 per cent lower than those estimated from the DM particles,
suggesting the presence of a small velocity bias. Using the best--fit
$\sigma_v$--$M_{200}$ relation obtained by \cite{2008ApJ...672..122E} from a
variety of N--body simulations, we obtain $M_{200}\simeq 2.7\times 10^{13}\hm$
for C1, and $M_{200}\simeq 1.9\times 10^{14}\hm$ for C2. These values are
similar to the true $M_{200}$ values (see Table 1), thus demonstrating that
virial equilibrium holds within $r_{200}$ in these proto--cluster regions. We
plot in Fig.~\ref{fi:Vel} the velocity distribution of all galaxies found in
projection within the same area shown in Fig.~\ref{fi:Maps_star}. The effect of
line-of-sight contamination from fore/background galaxies is evident for C1,
whose distribution shows an excess of galaxies with velocities $\magcir
1000\vel$.  For the C2 cluster, we have a non-negligible probability of
measuring a relative velocity as large as 2000 $\vel$, while this represents a
very unlike event for the C1 cluster. Therefore, even within the limited
statistics of available observations, the measured velocities for the flies of
the Spiderweb system are preferentially expected in a proto-cluster as massive
as C2.

\cite{2000A&A...361L..25P} measured redshifts for 15 Ly-$\alpha$ emitters in
the Spiderweb region and found a velocity dispersion $\sigma_v=900\pm 240 \vel$
(see also \citealt{2007A&A...461..823V}). However, the region sampled by these
Ly-$\alpha$ observations has a physical size of about $3\hm$, much larger than
the expected virialized region. In order to compare with these observational
data, we have computed $\sigma_v$ for the 50 most massive galaxies within
$3\hm$ from the BCGs of our simulated clusters. We find $\sigma_v= 331\vel$ for
C1 and $\sigma_v=683\vel$ for C2. This result indicates again that a
proto-cluster as massive as C2 is preferred with respect to the poorer C1
cluster. 

The too intense (with respect to observational estimates) star
formation that takes place in the C2 BCG suggests that AGN feedback
may have already partially quenched star formation in the Spiderweb
galaxy. Indeed, the Spiderweb galaxy has been originally identified as
a radio--galaxy with extended and distorted radio lobes
\citep{1998ApJ...504..139P}, consistent with the presence of a
``radio--mode'' AGN. Furthermore, \cite{2006ApJ...650..693N} carried
out an integral-field spectroscopic study of the Spiderweb complex and
found evidences for massive outflows of gas which are interpreted as
due to the action of AGN feedback.

\noindent{\bf Predicted X--ray properties:} Deep follow-up observations in the
X--ray band of a handful of clusters at $z>1$ are now pushing the study of
the thermo-- and chemo--dynamical properties of the ICM to large look-back
times. Although we have probably to wait for the next generation of X--ray
satellites to push these studies to $z\magcir 2$, it is interesting to make
predictions for the X--ray luminosity, temperature and level of metal enrichment
of proto--cluster regions, like the one traced by the Spiderweb galaxy.

Our simulations unambiguously predict that already at $z\simeq 2.1$, the
gravitational potential of proto--cluster regions are permeated by a hot
ICM. Indeed, for the C1 and C2 clusters we measure a temperature of about 2 and
5 keV, respectively (see Table 1). The X--ray luminosity of C1 is typical of a
nearby group, while that of C2 is comparable to the luminosity of a typical
bright nearby cluster.  At the redshift of the Spiderweb galaxy, these
luminosities correspond to fluxes in the [0.5--2] keV (observer frame) band of
about $1.0\times 10^{-15}$ for C1 and $1.5\times 10^{-14}\fl$ for C2, for the
cosmological model assumed in our simulations. A fairly high level of
enrichment is predicted already at $z\magcir 2$, extending to higher redshift
the results obtained out to $z\simeq 1.3$ with Chandra and XMM observations
\citep{2007A&A...462..429B,2008ApJS..174..117M}.

A relatively shallow Chandra observation, with an exposure of 40 ks,
has been used by \cite{2002ApJ...567..781C} to look for an extended
X--ray emission around the Spiderweb galaxy. These authors measured a
luminosity of $L\simeq 3.5\times 10^{44} \lum$ (rescaled to our
cosmology) in the [2--10]~keV band. They also placed an upper limit of
$1.7\times 10^{44} \lum$ for a relaxed cluster athmosphere. For C2, we
find a luminosity in the same band of about $6.8\times 10^{44}
\lum$. A luminosity of the hot gas surrounding the Spiderweb galaxy
lower than predicted for C2 can be interpreted as due to a significant
heating from the radio galaxy. Clearly, the relatively short exposure
time of this observation makes a precise determination of the X-ray
from a diffuse hot gas rather uncertain.

The flux predicted for C1 can in principle be reached with a sufficient long
exposure with the Chandra and XMM satellites.  However, surface brightness
dimming and the limited extension of the emission likely prevents it from being
detectable as an extended source using available X--ray telescopes. For C2, the
predicted flux is comparable to (or brighter than) that of clusters at $z>1$
observed with Chandra and XMM to study the thermal and chemical properties of
the ICM. A potential complication for the detection of an extended thermal
free--free emission from the hot gas permeating the Spiderweb complex could
come from non--thermal X--ray emission associated to star formation and/or
generated by the inverse--Compton scattering of a population of relativistic
electrons, associated to the radio galaxy, off the CMB photons
\citep[][]{2003MNRAS.341..729F}.

Although making a quantitative assessment of the observability of the hot ICM
surrounding the Spiderweb galaxy goes beyond the aims of this Letter, we point
out that such a target can be well suited to push the capabilities of the
present generation of X--ray satellites, while being optimal for large
collecting area telescopes of the next generation.

We note that our analysis is based on only two simulations. Clearly,
for a better assessment of the capability of current numerical models
in reproducing observations of proto--cluster structures it would be
ideal to carry out a statistical comparison by enlarging both the set
of simulated clusters and the number of observed proto--cluster
regions.

\section{Conclusions}
We have carried out an analysis of high--resolution hydrodynamical
simulations of two proto--cluster regions at $z\simeq 2.1$, with the
main purpose of comparing them with the observed properties of the
Spiderweb galaxy complex (M06, H08). These proto-clusters will form at
$z=0$ a relatively poor and a rich cluster of galaxies (C1 and C2
respectively; see Table 1). The comparison is aimed at demonstrating
to what extent predictions from simulations of proto--cluster regions
within a $\Lambda$CDM framework are able to reproduce the
observational properties of the Spiderweb complex, and to shed light
on the processes regulating star formation within the deepest
potential wells at $z\sim 2$. The main results of our analysis can be
summarised as follows.
\begin{itemize}
\item Both simulated proto--clusters are characterised by the presence of a
  dominant massive central galaxy, surrounded by less massive galaxies in the
  process of merging, qualitatively resembling the observed Spiderweb galaxy.
\item The star formation rate within the C1 proto--cluster ($\sim 600
  M_\odot{\rm yr}^{-1}$) is consistent with the observed one, while the
  corresponding value for C2 ($\simeq 1750 M_\odot{\rm yr}^{-1}$) is in excess
  with respect to observational measurements.
\item The BCG of C1 has an UV luminosity comparable to that of the Spiderweb
  galaxy, while the BCG of the C2 cluster is about one magnitude brighter.
\item The velocity dispersions of our simulated clusters within $r_{200}$ are
  consistent with virial equilibrium expectations. In addition, the velocities
  measured for the galaxies surrounding the Spiderweb system (M06,
  \citealt{2007A&A...461..823V}) are generally consistent with those of the C2
  cluster, while being much larger than those measured for C1.
\item The inter-galactic medium permeating the C1 and C2 proto--cluster regions
  has temperatures of about 2 keV and 5 keV respectively, and both clusters
  exhibit an enrichment level comparable to that observed in nearby clusters.
  The predicted X--ray fluxes from thermal bremsstrahlung make these objects
  potentially detectable as extended sources, and make them ideal targets to
  study the enrichment of the intra--cluster medium at unprecedented redshift,
  with X--ray telescopes of the next generation.
\end{itemize}

The emerging scenario is that the Spiderweb complex likely traces the
formation of a rich galaxy cluster, whose virial mass at $z=0$ is
$\sim 10^{15}\hm$. The excess of star formation found in our
simulations suggests that an AGN feedback might be necessary already
at $z\sim 2$ to quench star formation and regulate the structure of
the ``cool core'' \citep[see also][]{2006ApJ...650..693N}. The search
for proto-cluster regions at $z\magcir 2$ and their follow--up
observations with X--ray telescopes of the next generation will
contribute to fill the gap between $z\mincir 1$ studies of the thermo-
and chemo-dynamics of the ICM and the study of the IGM at $z>2$.

\section*{Acknowledgements}
We thank the referee, Brant Robertson, for insightful comments which helped us
to improve the presentation of the results. We are indebted to Volker Springel
for having provided us with the non--public version of \gadget. We thank
F. Fontanot, P. Monaco, R. Overzier, P. Rosati, L. Silva, and P. Tozzi for
useful discussions and comments on the paper. The simulations used in this
study have been carried out at the ``Centro Interuniversitario del Nord-Est per
il Calcolo Elettronico'' (CINECA, Bologna), with CPU time assigned through an
INAF--CINECA grant. This work has been partially supported by the INFN PD-51
grant, by the INAF-PRIN06 Grant and by the ASI-COFIS Theory grant.

\bibliographystyle{mn2e}
\bibliography{master}

\begin{thebibliography}{}

\bibitem[\protect\citeauthoryear{{Balestra}, {Tozzi}, {Ettori}, {Rosati},
  {Borgani}, {Norman} \& {Viola}}{{Balestra}
  et~al.}{2007}]{2007A&A...462..429B}
{Balestra} I.,  {Tozzi} P.,  {Ettori} S.,  {Rosati} P.,  {Borgani} S.,
  {Norman} V.~M.~C.,    {Viola} M.,  2007, \aap, 462, 429

\bibitem[\protect\citeauthoryear{{Borgani}, {Dolag}, {Murante}, {Cheng},
  {Springel}, {Diaferio}, {Moscardini}, {Tormen}, {Tornatore} \&
  {Tozzi}}{{Borgani} et~al.}{2006}]{2006MNRAS.367.1641B}
{Borgani} S.,  {Dolag} K.,  {Murante} G.,  {Cheng} L.-M.,  {Springel} V.,
  {Diaferio} A.,  {Moscardini} L.,  {Tormen} G.,  {Tornatore} L.,    {Tozzi}
  P.,  2006, \mnras, 367, 1641

\bibitem[\protect\citeauthoryear{{Borgani}, {Murante}, {Springel}, {Diaferio},
  {Dolag}, {Moscardini}, {Tormen}, {Tornatore} \& {Tozzi}}{{Borgani}
  et~al.}{2004}]{2004MNRAS.348.1078B}
{Borgani} S.,  {Murante} G.,  {Springel} V.,  {Diaferio} A.,  {Dolag} K.,
  {Moscardini} L.,  {Tormen} G.,  {Tornatore} L.,    {Tozzi} P.,  2004, \mnras,
  348, 1078

\bibitem[\protect\citeauthoryear{{Bruzual} \& {Charlot}}{{Bruzual} \&
  {Charlot}}{2003}]{2003MNRAS.344.1000B}
{Bruzual} G.,  {Charlot} S.,  2003, \mnras, 344, 1000

\bibitem[\protect\citeauthoryear{{Carilli}, {Harris}, {Pentericci},
  {R{\"o}ttgering}, {Miley}, {Kurk} \& {van Breugel}}{{Carilli}
  et~al.}{2002}]{2002ApJ...567..781C}
{Carilli} C.~L.,  {Harris} D.~E.,  {Pentericci} L.,  {R{\"o}ttgering} H.~J.~A.,
   {Miley} G.~K.,  {Kurk} J.~D.,    {van Breugel} W.,  2002, \apj, 567, 781

\bibitem[\protect\citeauthoryear{{Charlot} \& {Fall}}{{Charlot} \&
  {Fall}}{2000}]{2000ApJ...539..718C}
{Charlot} S.,  {Fall} S.~M.,  2000, \apj, 539, 718

\bibitem[\protect\citeauthoryear{{De Lucia} \& {Blaizot}}{{De Lucia} \&
  {Blaizot}}{2007}]{2007MNRAS.375....2D}
{De Lucia} G.,  {Blaizot} J.,  2007, \mnras, 375, 2

\bibitem[\protect\citeauthoryear{{Dolag}, {Borgani}, {Murante} \&
  {Springel}}{{Dolag} et~al.}{2008}]{2008arXiv0808.3401D}
{Dolag} K.,  {Borgani} S.,  {Murante} G.,    {Springel} V.,  2008, ArXiv
  e-prints

\bibitem[\protect\citeauthoryear{{Eisenhardt}, {Brodwin}, {Gonzalez},
  {Stanford}, {Stern}, {Barmby}, {Brown}, {Dawson}, {Dey}, {Doi}, {Galametz},
  {Jannuzi}, {Kochanek}, {Meyers}, {Morokuma} \& {Moustakas}}{{Eisenhardt}
  et~al.}{2008}]{2008arXiv0804.4798E}
{Eisenhardt} P.~R.~M.,  {Brodwin} M.,  {Gonzalez} A.~H.,  {Stanford} S.~A.,
  {Stern} D.,  {Barmby} P.,  {Brown} M.~J.~I.,  {Dawson} K.,  {Dey} A.,  {Doi}
  M.,  {Galametz} A.,  {Jannuzi} B.~T.,  {Kochanek} C.~S.,  {Meyers} J.,
  {Morokuma} T.,    {Moustakas} L.~A.,  2008, ArXiv e-prints, 804

\bibitem[\protect\citeauthoryear{{Evrard}, {Bialek}, {Busha}, {White}, {Habib},
  {Heitmann}, {Warren}, {Rasia}, {Tormen}, {Moscardini}, {Power}, {Jenkins},
  {Gao}, {Frenk}, {Springel}, {White} \& {Diemand}}{{Evrard}
  et~al.}{2008}]{2008ApJ...672..122E}
{Evrard} A.~E.,  {Bialek} J.,  {Busha} M.,  {White} M.,  {Habib} S.,
  {Heitmann} K.,  {Warren} M.,  {Rasia} E.,  {Tormen} G.,  {Moscardini} L.,
  {Power} C.,  {Jenkins} A.~R.,  {Gao} L.,  {Frenk} C.~S.,  {Springel} V.,
  {White} S.~D.~M.,    {Diemand} J.,  2008, \apj, 672, 122

\bibitem[\protect\citeauthoryear{{Fabian}, {Sanders}, {Crawford} \&
  {Ettori}}{{Fabian} et~al.}{2003}]{2003MNRAS.341..729F}
{Fabian} A.~C.,  {Sanders} J.~S.,  {Crawford} C.~S.,    {Ettori} S.,  2003,
  \mnras, 341, 729

\bibitem[\protect\citeauthoryear{{Fabjan}, {Tornatore}, {Borgani}, {Saro} \&
  {Dolag}}{{Fabjan} et~al.}{2008}]{2008MNRAS.386.1265F}
{Fabjan} D.,  {Tornatore} L.,  {Borgani} S.,  {Saro} A.,    {Dolag} K.,  2008,
  \mnras, 386, 1265

\bibitem[\protect\citeauthoryear{{Fontanot}, {Somerville}, {Silva}, {Monaco} \&
  {Skibba}}{{Fontanot} et~al.}{2008}]{Fontanot08}
{Fontanot} F.,  {Somerville} R.S., {Silva} L., {Monaco} P., {Skibba} R., 2008,
  \mnras, submitted

\bibitem[\protect\citeauthoryear{{Kong}, {Charlot}, {Brinchmann} \&
  {Fall}}{{Kong} et~al.}{2004}]{2004MNRAS.349..769K}
{Kong} X.,  {Charlot} S.,  {Brinchmann} J.,    {Fall} S.~M.,  2004, \mnras,
  349, 769

\bibitem[\protect\citeauthoryear{{Maughan}, {Jones}, {Forman} \& {Van
  Speybroeck}}{{Maughan} et~al.}{2008}]{2008ApJS..174..117M}
{Maughan} B.~J.,  {Jones} C.,  {Forman} W.,    {Van Speybroeck} L.,  2008,
  \apjs, 174, 117

\bibitem[\protect\citeauthoryear{{Miley}, {Overzier}, {Zirm}, {Ford}, {Kurk},
  {Pentericci}, {Blakeslee}, {Franx}, {Illingworth}, {Postman}, {Rosati},
  {R{\"o}ttgering}, {Venemans} \& {Helder}}{{Miley}
  et~al.}{2006}]{2006ApJ...650L..29M}
{Miley} G.~K.,  {Overzier} R.~A.,  {Zirm} A.~W.,  {Ford} H.~C.,  {Kurk} J.,
  {Pentericci} L.,  {Blakeslee} J.~P.,  {Franx} M.,  {Illingworth} G.~D.,
  {Postman} M.,  {Rosati} P.,  {R{\"o}ttgering} H.~J.~A.,  {Venemans} B.~P.,
  {Helder} E.,  2006, \apjl, 650, L29 (M06)

\bibitem[\protect\citeauthoryear{{Mullis}, {Rosati}, {Lamer}, {B{\"o}hringer},
  {Schwope}, {Schuecker} \& {Fassbender}}{{Mullis}
  et~al.}{2005}]{2005ApJ...623L..85M}
{Mullis} C.~R.,  {Rosati} P.,  {Lamer} G.,  {B{\"o}hringer} H.,  {Schwope} A.,
  {Schuecker} P.,    {Fassbender} R.,  2005, \apjl, 623, L85

\bibitem[\protect\citeauthoryear{{Murante}, {Giovalli}, {Gerhard}, {Arnaboldi},
  {Borgani} \& {Dolag}}{{Murante} et~al.}{2007}]{2007MNRAS.377....2M}
{Murante} G.,  {Giovalli} M.,  {Gerhard} O.,  {Arnaboldi} M.,  {Borgani} S.,
  {Dolag} K.,  2007, \mnras, 377, 2

\bibitem[\protect\citeauthoryear{{Nesvadba}, {Lehnert}, {Eisenhauer},
  {Gilbert}, {Tecza} \& {Abuter}}{{Nesvadba}
  et~al.}{2006}]{2006ApJ...650..693N}
{Nesvadba} N.~P.~H.,  {Lehnert} M.~D.,  {Eisenhauer} F.,  {Gilbert} A.,
  {Tecza} M.,    {Abuter} R.,  2006, \apj, 650, 693

\bibitem[\protect\citeauthoryear{{Overzier}, {Bouwens}, {Cross}, {Venemans},
  {Miley}, {Zirm}, {Ben{\'{\i}}tez}, {Blakeslee}, {Coe}, {Demarco} \&
  {Ford}}{{Overzier} et~al.}{2008}]{2008ApJ...673..143O}
{Overzier} R.~A.,  {Bouwens} R.~J.,  {Cross} N.~J.~G.,  {Venemans} B.~P.,
  {Miley} G.~K.,  {Zirm} A.~W.,  {Ben{\'{\i}}tez} N.,  {Blakeslee} J.~P.,
  {Coe} D.,  {Demarco} R.,    {Ford} 2008, \apj, 673, 143

\bibitem[\protect\citeauthoryear{{Pentericci}, {Kurk}, {R{\"o}ttgering},
  {Miley}, {van Breugel}, {Carilli}, {Ford}, {Heckman}, {McCarthy} \&
  {Moorwood}}{{Pentericci} et~al.}{2000}]{2000A&A...361L..25P}
{Pentericci} L.,  {Kurk} J.~D.,  {R{\"o}ttgering} H.~J.~A.,  {Miley} G.~K.,
  {van Breugel} W.,  {Carilli} C.~L.,  {Ford} H.,  {Heckman} T.,  {McCarthy}
  P.,    {Moorwood} A.,  2000, \aap, 361, L25

\bibitem[\protect\citeauthoryear{{Pentericci}, {Roettgering}, {Miley},
  {Spinrad}, {McCarthy}, {van Breugel} \& {Macchetto}}{{Pentericci}
  et~al.}{1998}]{1998ApJ...504..139P}
{Pentericci} L.,  {Roettgering} H.~J.~A.,  {Miley} G.~K.,  {Spinrad} H.,
  {McCarthy} P.~J.,  {van Breugel} W.~J.~M.,    {Macchetto} F.,  1998, \apj,
  504, 139

\bibitem[\protect\citeauthoryear{{Pierini}, {Zibetti}, {Braglia},
  {B{\"o}hringer}, {Finoguenov}, {Lynam} \& {Zhang}}{{Pierini}
  et~al.}{2008}]{2008A&A...483..727P}
{Pierini} D.,  {Zibetti} S.,  {Braglia} F.,  {B{\"o}hringer} H.,  {Finoguenov}
  A.,  {Lynam} P.~D.,    {Zhang} Y.-Y.,  2008, \aap, 483, 727

\bibitem[\protect\citeauthoryear{{Romeo}, {Napolitano}, {Covone},
  {Sommer-Larsen}, {Antonuccio-Delogu} \& {Capaccioli}}{{Romeo}
  et~al.}{2008}]{2008MNRAS.tmp..793R}
{Romeo} A.~D.,  {Napolitano} N.~R.,  {Covone} G.,  {Sommer-Larsen} J.,
  {Antonuccio-Delogu} V.,    {Capaccioli} M.,  2008, \mnras, pp 793--+

\bibitem[\protect\citeauthoryear{{Salpeter}}{{Salpeter}}{1955}]{1955ApJ...121.%
.161S}
{Salpeter} E.~E.,  1955, \apj, 121, 161

\bibitem[\protect\citeauthoryear{{Saro}, {Borgani}, {Tornatore}, {Dolag},
  {Murante}, {Biviano}, {Calura} \& {Charlot}}{{Saro}
  et~al.}{2006}]{2006MNRAS.373..397S}
{Saro} A.,  {Borgani} S.,  {Tornatore} L.,  {Dolag} K.,  {Murante} G.,
  {Biviano} A.,  {Calura} F.,    {Charlot} S.,  2006, \mnras, 373, 397

\bibitem[\protect\citeauthoryear{{Springel}}{{Springel}}{2005}]{2005MNRAS.364.%
1105S}
{Springel} V.,  2005, \mnras, 364, 1105

\bibitem[\protect\citeauthoryear{{Springel} \& {Hernquist}}{{Springel} \&
  {Hernquist}}{2003}]{2003MNRAS.339..289S}
{Springel} V.,  {Hernquist} L.,  2003, \mnras, 339, 289

\bibitem[\protect\citeauthoryear{{Stadel}}{{Stadel}}{2001}]{2001PhDT........21%
S}
{Stadel} J.~G.,  2001, Ph.D.~Thesis

\bibitem[\protect\citeauthoryear{{Stanford}, {Eisenhardt}, {Brodwin},
  {Gonzalez}, {Stern}, {Jannuzi}, {Dey}, {Brown}, {McKenzie} \&
  {Elston}}{{Stanford} et~al.}{2005}]{2005ApJ...634L.129S}
{Stanford} S.~A.,  {Eisenhardt} P.~R.,  {Brodwin} M.,  {Gonzalez} A.~H.,
  {Stern} D.,  {Jannuzi} B.~T.,  {Dey} A.,  {Brown} M.~J.~I.,  {McKenzie} E.,
   {Elston} R.,  2005, \apjl, 634, L129

\bibitem[\protect\citeauthoryear{{Tormen}, {Bouchet} \& {White}}{{Tormen}
  et~al.}{1997}]{1997MNRAS.286..865T}
{Tormen} G.,  {Bouchet} F.~R.,    {White} S.~D.~M.,  1997, \mnras, 286, 865

\bibitem[\protect\citeauthoryear{{Tornatore}, {Borgani}, {Dolag} \&
  {Matteucci}}{{Tornatore} et~al.}{2007}]{2007MNRAS.382.1050T}
{Tornatore} L.,  {Borgani} S.,  {Dolag} K.,    {Matteucci} F.,  2007, \mnras,
  382, 1050

\bibitem[\protect\citeauthoryear{{Venemans}, {R{\"o}ttgering}, {Miley}, {van
  Breugel}, {de Breuck}, {Kurk}, {Pentericci}, {Stanford}, {Overzier}, {Croft}
  \& {Ford}}{{Venemans} et~al.}{2007}]{2007A&A...461..823V}
{Venemans} B.~P.,  {R{\"o}ttgering} H.~J.~A.,  {Miley} G.~K.,  {van Breugel}
  W.~J.~M.,  {de Breuck} C.,  {Kurk} J.~D.,  {Pentericci} L.,  {Stanford}
  S.~A.,  {Overzier} R.~A.,  {Croft} S.,    {Ford} H.,  2007, \aap, 461, 823

\bibitem[\protect\citeauthoryear{{Yoshida}, {Sheth} \& {Diaferio}}{{Yoshida}
  et~al.}{2001}]{2001MNRAS.328..669Y}
{Yoshida} N.,  {Sheth} R.~K.,    {Diaferio} A.,  2001, \mnras, 328, 669

\bibitem[\protect\citeauthoryear{{Zibetti}, {White}, {Schneider} \&
  {Brinkmann}}{{Zibetti} et~al.}{2005}]{2005MNRAS.358..949Z}
{Zibetti} S.,  {White} S.~D.~M.,  {Schneider} D.~P.,    {Brinkmann} J.,  2005,
  \mnras, 358, 949

\end{thebibliography}

\end{document}